\documentclass[aps,12pt,final,notitlepage,oneside,onecolumn
,nobibnotes,nofootinbib,superscriptaddress,showpacs
,tightenlines,amssymb] 
{revtex4}

\usepackage[dvips]{graphicx}

\newcommand{\beq}{\begin{equation}}
\newcommand{\eeq}{\end{equation}}
\newcommand{\bea}{\begin{eqnarray}}
\newcommand{\eea}{\end{eqnarray}}

\def\eps{\varepsilon}
\def\sgm{\Sigma^-}
\def\sg{\Sigma}
\def\la{\Lambda}
\def\non{\nonumber}
\def\ron{\rho_N}
\def\rol{\rho_\Lambda}
\def\ros{\rho_\Sigma}
\def\ms{M_\odot}

\begin{document}

\title{
Hadron-Quark Phase Transitions in Hyperon Stars}

\author{N. Yasutake}
\affiliation{
Division of Theoretical Astronomy, 
National Astronomical Observatory of Japan, 
2-21-1 Osawa, Mitaka, Tokyo 181-8588, Japan}

\author{G.F. Burgio} \author{H.-J. Schulze} 
\affiliation{
INFN, Sezione di Catania, 
Via Santa Sofia 64, I-95123 Catania, Italy}

\begin{abstract}
We compare the Gibbs and Maxwell constructions
for the hadron-quark phase transition
in neutron and protoneutron stars,
including interacting hyperons in the confined phase.
We find that the hyperon populations are suppressed
and that neutrino trapping shifts the onset of the phase transition.
The effects on the (proto)neutron star maximum mass are explored.
\end{abstract}


\maketitle

\section{Introduction}

Some recent lattice QCD calculations have improved our knowledge on the nature
of the nucleon-hyperon (NY) and hyperon-hyperon (YY) interactions 
\cite{nemura09}. 
Moreover, an intense experimental activity aimed at exploring hypernuclei has 
started at the J-PARC facility, which will hopefully clarify in the near future
some unresolved questions on the hyperon interactions.
This will help to improve the theoretical modeling of neutron star (NS) 
interiors, 
where hyperons are predicted to be present in large fractions, 
and many aspects of the equation of state (EOS) could be clarified.

There are essentially two microscopic approaches to derive the properties 
and the EOS of dense matter starting from the bare baryon-baryon interactions,
namely the Brueckner-Hartree-Fock (BHF) theory \cite{book} 
and the variational method \cite{apr}.
Including the hyperon degrees of freedom, the EOS becomes very soft,
and in the BHF approach the maximum mass for a NS lies below 
currently observed values of about 1.7 $\ms$ \cite{obs}.
To overcome this problem, some possibilities have been suggested:
i) The repulsive effects of three-body forces (TBF) among nucleons and hyperons, 
which may stiffen the EOS \cite{nishizaki02};
ii) Additional repulsion coming from hyperon-hyperon interactions 
(but no experimental data are available so far);
iii) The hadron-quark phase transition \cite{burgio02,maieron04}, 
which could lead to a partial suppression of the hyperon population, 
thus stiffening the EOS. 
This conclusion needs to be further explored, 
since at present there are many uncertainties regarding the quark matter EOS. 

The scope of this work is to present results on the composition and structure 
of neutron and protoneutron stars (PNS), 
which are produced in the aftermath of
successful supernova (SN) explosions 
of very massive stars ($M\gtrsim8\;\ms$).
These objects can reach temperatures of the order of 30--40 MeV in their interiors,
and are characterized by a temporary neutrino trapping
with a conserved lepton fraction $Y_e \approx$ 0.4 \cite{burr,pons},
lasting some seconds. 
Both thermal effects and neutrino trapping may result in observable 
consequences of the neutrino signature from a supernova, 
and may also play an important role in determining whether a SN ultimately 
produces a cold NS or a black hole. 
In this paper, we focus on the hadron-quark phase transition, 
with the explicit inclusion of interacting hyperons in the confined phase.
For hadronic matter we adopt the BHF microscopic EOS extended to finite temperature, 
and use the MIT bag model in the quark phase.

This paper is organized as follows.
In Section II we illustrate the Brueckner-Bethe-Goldstone (BBG)
many-body theory including hyperons at finite temperature.
Sections III and IV briefly review the MIT bag model and the treatment of the
hadron-quark mixed phase, respectively.
In Section V we discuss our results regarding the structure of NSs and PNSs, 
in particular their maximum mass. 
Final conclusions are drawn in Section VI.

\section{Hadron phase: The BBG theory at finite temperature}

In the recent years, the BBG perturbative theory \cite{book}
has been extended in a fully microscopic way to the
finite-temperature case \cite{big1},
according to the formalism developed by Bloch and De Dominicis \cite{bloch}. 
In this approach the essential ingredient is the two-body scattering matrix $G$,
which in the finite-temperature extension 
is determined by solving numerically the Bethe-Goldstone equation,
written in operatorial form as 
\beq
  G_{ab}[W] = V_{ab} + 
  \sum_c \sum_{p,p'}
  V_{ac} \big|pp'\big\rangle
  { Q_c \over W - E_c +i\eps}
  \big\langle pp'\big| G_{cb}[W] \:,
\label{e:g}
\eeq
where the indices $a,b,c$ indicate pairs of baryons
and the Pauli operator $Q$ and energy $E$
determine the propagation of intermediate baryon pairs.
In a given baryon-baryon channel $c=(12)$ one has
\beq
   Q_{(12)} = [1-n_1(k_1)][1-n_2(k_2)] \:,
\eeq
\beq
  E_{(12)} = m_1 + m_2 + e_1(k_1) + e_2(k_2)
\label{e:e}
\eeq
with the single-particle energy
$e_i(k) = k^2\!/2m_i + U_i(k)$, 
the Fermi distribution
$n_i(k)=\big({\rm e}^{[e_i(k) - \tilde{\mu}_i]/T} + 1 \big)^{-1}$,
the starting energy $W$,
and the two-body interaction (bare potential)
$V$ as fundamental input.
The various single-particle potentials within the continuous choice
are given by
\beq
  U_1(k_1) = {\rm Re}\!\!\!\! 
  \sum_{2=n,p,\la,\sg}\sum_{k_2} n(k_2)
  \big\langle k_1 k_2 \big| G_{(12)(12)}[E_{(12)}]
  \big| k_1 k_2 \big\rangle_A \:,
\label{e:u}
\eeq
where $k_i$ generally denote momentum and spin.

We choose the Argonne $V_{18}$ nucleon-nucleon potential \cite{wiringa} 
as two-body interaction $V$, 
supplemented by TBF among nucleons, 
in order to reproduce correctly the nuclear matter saturation point 
$\rho_0 \approx 0.17~\mathrm{fm}^{-3}$, $E/A\approx -16$ MeV. 
As TBF, we use the phenomenological Urbana model \cite{schi}, 
which is actually reduced to a density-dependent two-body force 
by averaging over the position of the third particle \cite{big1,bbb}. 

Recently, the BHF approach has been extended by including consistently 
interacting hyperons at finite temperature \cite{bsa}. 
In the hyperonic sector 
we employed the Nijmegen soft-core NY potentials NSC89~\cite{nsc89},
fitted to the available experimental NY scattering data.
Since at zero temperature only $\la$ and $\sgm$ hyperons
appear in the neutron star matter up to very large densities \cite{hypns}, 
we restrict also the present study to these two hyperon species.
Therefore, for fixed partial densities 
$\rho_i = \sum_k n_i(k)$, $(i=n,p,\la,\sgm)$
and temperature $T$,  
we solve self-consistently Eqs.~(\ref{e:g}) and (\ref{e:u}) 
and calculate then the free energy density, 
which has the following simplified expression
\begin{equation}
 f = \sum_i \left[ \sum_{k} n_i(k)
 \left( {k^2\over 2m_i} + {1\over 2}U_i(k) \right) - Ts_i \right] \:,
\label{e:f}
\end{equation}
where
\begin{equation}
 s_i = - \sum_{k} \Big( n_i(k) \ln n_i(k) + [1-n_i(k)] \ln [1-n_i(k)] \Big)
\end{equation}
is the entropy density for component $i$ treated as a free gas with
spectrum $e_i(k)$. 
All thermodynamic quantities of interest can then be computed 
from the free energy density, Eq.~(\ref{e:f}); 
namely, the chemical potentials $\mu_i$, pressure $p$, 
entropy density $s$, and internal energy density $\eps$ read as
\bea
 \mu_i &=& {{\partial f}\over{\partial \rho_i}} \:,
\\
 p &=& \rho^2 {{\partial (f/\rho)}\over{\partial \rho} }  
 = \sum_i \mu_i \rho_i - f  \:,
\label{e:p}
\\
 s &=& -{{\partial f}\over{\partial T}} \:,
\\
 \eps &=& f + Ts \:,
\label{e:eps}
\eea  
where $\rho=\sum_i\rho_i$ is the baryon number density.

However, due to the large number of degrees of freedom 
(4 partial densities + temperature), 
it is necessary to introduce
some approximations in order to speed up the calculations. 
We adopt the so-called Frozen Correlations Approximation (FCA),
assuming the correlations at $T\neq 0$ to be essentially the same as at $T=0$. 
This means that the single-particle potential $U_i(k)$ for the component $i$ 
can be approximated by the one calculated at $T=0$. 
Furthermore, we fit the numerical results 
by a sufficiently accurate analytical parametrization.
We find that the following functional form provides an excellent 
parametrization of the numerical data for the free energy density
in the required ranges of nucleon density 
($0.1\;{\rm fm}^{-3} \lesssim \rho_N \lesssim 0.8\;{\rm fm}^{-3}$),
hyperon fractions
($0 \leq \rol/\rho_N \leq 0.9$, $0 \leq \ros/\rho_N \leq 0.5$),
and temperature ($0\;{\rm MeV} \leq T \leq 50\;{\rm MeV}$):
\bea
 && f(\rho_n,\rho_p,\rol,\ros,T) = F_N \ron
\non\\&&\quad  
 + \left(F_\la+F_{\la\la}+F_{\la\sg}\right)\rol + {C \over 2m_\la M_\la}\rol^{5/3}
\non\\&&\quad 
 + \left(F_\sg+F_{\sg\sg}+F_{\sg\la}\right)\ros + {C \over 2m_\sg M_\sg}\ros^{5/3}
\label{e:fps}
\eea
with the parametrizations at zero temperature:
\bea
  F_N &=& (1-\beta) \left( a_0 \ron + b_0 \ron^{c_0} \right) 
  + \beta  \left( a_1 \ron + b_1 \ron^{c_1} \right) 
\:,\\
  F_Y &=& (a_Y^0 + a_Y^1 x + a_Y^2 x^2) \ron 
  + (b_Y^0 + b_Y^1 x + b_Y^2 x^2) \ron^{c_Y} 
\:,\qquad\\
  F_{YY'} &=& a_{YY'} \ron^{c_{YY'}} \rho_{Y'}^{d_{YY'}} 
\:,\\
  M_Y &=& 1 + \left( c_Y^0 + c_Y^1 x \right) \ron 
\:,\eea
where
$\ron=\rho_n+\rho_p$;
$x=\rho_p/\ron$;
$\beta = (1-2x)^2$;
$C = (3/5)(3\pi^2)^{2/3}\approx 5.742$;
and
$Y,Y'=\la,\sgm$.
At finite temperature the expressions are extended as follows:
\bea
 F_N &=& F_N(T=0)
\non\\&+&
 \left[
 \tilde{a}_0 t^2 \rho_N + (\tilde{d}_0 t^2 + \tilde{e}_0 t^3)\ln(\rho_N) 
 + \tilde{f}_0 t^2/\rho_N 
 \right](1-\beta)
\non\\&+&
 \left[
 \tilde{a}_1 t^2 \rho_N + (\tilde{d}_1 t^2 + \tilde{e}_1 t^3)\ln(\rho_N) 
 + \tilde{f}_1 t^2/\rho_N 
 \right] \beta
\:,
\\
 F_Y &=& F_Y(T=0)
\non\\&+&
 (\tilde{d}_Y t^2 + \tilde{e}_Y t^1)\ln(\rho_N) + \tilde{f}_Y t^2/\rho_N 
 + \tilde{g}_Y t^2 \ln(\rho_Y)
\:,\qquad
\label{e:fitfs}
\\
 M_Y &=& M_Y(T=0) 
 + \tilde{b}_Y t^2 \ron^{\tilde{c}_Y} \:,
\label{e:fpsend}
\eea
where $t=T/(100\,\mathrm{MeV})$ and $f$ and $\rho_i$ are given in
MeV\,fm$^{-3}$ and fm$^{-3}$, respectively,
(and $m_{\la,\sg}$ in MeV$^{-1}$fm$^{-2}$).

\renewcommand{\arraystretch}{0.6}
\begin{table}[t]
\caption{Fit parameters for the free energy density, 
Eqs.~(\ref{e:fps}-\ref{e:fpsend}).}
\begin{tabular}{l|rrrrrrr}
\hline
\hline
 $a_0,b_0,c_0,a_1,b_1,c_1$              
 & -286.6  & 397.2  & 1.39 & 88.1 & 207.7 & 2.50 & \\
 $a_\la^0,a_\la^1,a_\la^2,b_\la^0,b_\la^1,b_\la^2,c_\la$  
 & -403  & 688  & -943 & 659 & -1273 & 1761 & 1.72 \\
 $a_\sg^0,a_\sg^1,a_\sg^2,b_\sg^0,b_\sg^1,b_\sg^2,c_\sg$  
 & -114  & 0  & 0 & 291 & 0 & 0 & 1.63 \\
 $a_{\la\la},c_{\la\la},d_{\la\la}$
 & 136 & 0.51 & 0.93 & & & & \\ 
 $a_{\la\sg},c_{\la\sg},d_{\la\sg}$
 & 0 & 0 & 0 & & & & \\ 
 $a_{\sg\sg},c_{\sg\sg},d_{\sg\sg}$
 & 0 & 0 & 0 & & & & \\ 
 $a_{\sg\la},c_{\sg\la},d_{\sg\la}$
 & 89 & 0.33 & 0.81 & & & & \\ 
 $c_\la^0,c_\la^1,c_\sg^0,c_\sg^1$ \strut             
 & 0.22  & -0.38  & -0.59 & -0.22 &  & & \\
\hline
 $\tilde{a}_0,\tilde{d}_0,\tilde{e}_0,\tilde{f}_0$
 & -202.0  & 396.9  & -190.6 & 35.2 & & & \\
 $\tilde{a}_1,\tilde{d}_1,\tilde{e}_1,\tilde{f}_1$
 & -138.0  & 308.4  & -109.3 & 31.2 & & & \\
 $\tilde{d}_\la,\tilde{e}_\la,\tilde{f}_\la,\tilde{g}_\la,\tilde{b}_\la,\tilde{c}_\la$
 & 92.3  & 29.3  & 39.4 & 152.3 & 4.78 & 3.95 & \\
 $\tilde{d}_\sg,\tilde{e}_\sg,\tilde{f}_\sg,\tilde{g}_\sg,\tilde{b}_\sg,\tilde{c}_\sg$ \strut
 & 89.2  & 61.0  & 63.6 & 186.8 & 1.13 & 3.30 & \\
\hline
\hline
\end{tabular}
\label{t:trans}
\end{table}

Technically,
these parametrizations were obtained by performing about $10^3$ BHF calculations
at zero temperature 
in the $(\rho_n,\rho_p,\rol,\ros)$--space
and then using the FCA to generate finite-temperature results,
increasing by about one order of magnitude the number 
of ``data'' points $f(\rho_n,\rho_p,\rol,\ros,T)$. 
The optimal values of the fit parameters,
listed in Table I, 
were then determined hierarchically for
cold nuclear matter, cold hypernuclear matter,
hot nuclear matter, hot hypernuclear matter,
so that the fits are optimized also in the more constrained cases \cite{bsa}.

\section{Quark phase: The MIT bag model}

For the quark phase, we adopt the MIT bag model 
involving $u$, $d$ and $s$ quarks \cite{burgio02,maieron04,nic06}.
Probably, this model is too simple to describe quark matter in a realistic way, 
and we plan to adopt more sophisticated models in the future 
\cite{njl,huan09,yasutake09a}.
We assume massless $u$ and $d$ quarks, 
$s$ quarks with a current mass of $m_s=150$ MeV, 
and a density-dependent 
(but temperature-independent)
bag constant, 
\begin{eqnarray}
 B(\rho) = B_\infty +(B_0 - B_\infty) \exp\!\Big[-\beta \Big( \frac{\rho}{\rho_0} 
 \Big )^2\Big ]
\label{eq:01}
\end{eqnarray}
with $B_\infty = 50$ MeV\,fm$^{-3}$, $B_0=400$ MeV\,fm$^{-3}$, and $\beta=0.17$. 

This approach has been proposed in Ref.~\cite{burgio02}
on the basis of experimental results on the formation of a quark-gluon plasma
obtained at the CERN SPS.
The above choice of the parameters allows the symmetric nuclear matter
to be in the pure hadronic phase at low densities, 
and in the quark phase at large densities, 
while the transition density is taken as a parameter. 
Several possible choices of the parameters have been explored in \cite{burgio02}, 
and all give a NS maximum mass in a relatively narrow interval, 
$1.4\;\ms \lesssim M_{\rm max} \lesssim 1.7\;\ms$. 
For a more extensive discussion of this topic, 
the reader is referred to \cite{nic06}, 
where details regarding the MIT bag model at finite temperature are also given, 
which will not be repeated here.

\section{Mixed phase}

With this parametrization of the density dependence of $B$ we now
consider the hadron-quark phase transition in (proto)neutron stars.
We calculate the EOS of a conventional neutron star as composed of a 
chemically equilibrated and charge-neutral mixture of 
nucleons, hyperons, and leptons. 
We do not take into account anti-particles and muons in this paper, 
since their effects on the EOS are very small. 

We compare numerical results 
obtained using the Maxwell and the Gibbs constructions for the phase transition.
We remind that in the Maxwell construction both phases
are charge neutral and the conditions of mechanical and chemical equilibrium 
determine a coexistence phase, which is described by a constant pressure 
plateau typical of a liquid-vapor phase transition. 
On the other hand, in the Gibbs construction \cite{glen} 
both the hadron and the quark phase are allowed to be separately charged, 
still preserving the total charge neutrality. 
This implies that neutron star matter can be treated as a two-component system, 
and therefore can be parametrized by two chemical potentials.
Usually one chooses the pair ($\mu_e, \mu_n$), 
i.e., electron and baryon chemical potential. 
The pressure is the same in the two phases to ensure mechanical stability, 
while the chemical potentials of the different species are related to each other,
satisfying chemical and beta stability. 
As a consequence,
the pressure turns out to be a monotonically increasing function of the density,
at variance with the Maxwell construction.
We note that our Gibbs treatment is the zero surface tension limit 
in the calculations including finite-size effects \cite{maruyama07,yasutake09}.

The Gibbs conditions for chemical and mechanical equilibrium 
at finite temperature between both phases read
\begin{eqnarray}
\mu_u + \mu_e - \mu_{\nu{_e}} & = & \mu_d = \mu_s \:, 
\\
\mu_p + \mu_e - \mu_{\nu{_e}} & = & \mu_n = \mu_\Lambda = \mu_u + 2\mu_d \:,
\\
\mu_{\Sigma^-} + \mu_p & = & 2\mu_n \:,
\\
p_H(\mu_e,\mu_n,T) & = &  p_Q(\mu_e,\mu_n,T) = p_M \:. 
\label{e:mp}
\end{eqnarray}
From these equations one can calculate the equilibrium chemical potentials 
of the mixed phase corresponding to the intersection of the two surfaces 
representing the hadron and the quark phase, 
which allows one to calculate the charge densities $\rho_c^H$ and $\rho_c^Q$ 
and therefore the volume fraction $\chi$ occupied 
by quark matter in the mixed phase, i.e.,
\begin{equation}
 \chi \rho_c^Q + (1 - \chi) \rho_c^H = 0 \:.
\label{e:chi}
\end{equation}
From this, the energy density $\epsilon_M$ and the baryon density $\rho_M$ 
of the mixed phase can be determined as
\begin{eqnarray}
 \epsilon_M &=& \chi \epsilon_Q + (1 - \chi)\epsilon_H \:, 
\\
 \rho_M &=& \chi \rho_Q + (1 - \chi)\rho_H \:. 
\label{e:mp1}
\end{eqnarray}


\section{Results and discussion}

\begin{figure}[t] 
\includegraphics[clip,scale=2]{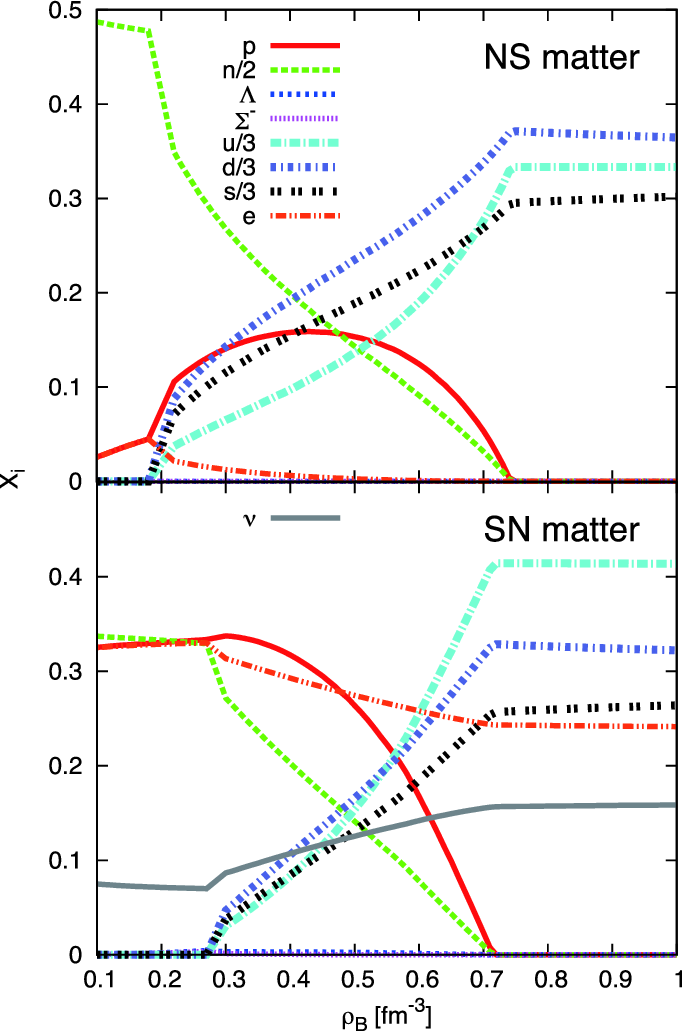}
\caption{
Particle fractions $x_i=\rho_i/\rho_B$ 
as a function of the baryon density $\rho_B$
for NS (upper panel) and SN matter (lower panel). 
The mixed phase is calculated by the Gibbs conditions.} 
\label{fig:01}
\end{figure}

In Fig.~\ref{fig:01} we show the particle fractions obtained by applying the 
Gibbs construction to NS matter
($T=0$ MeV and $x_{\nu_e}=0$, upper panel), 
and to SN matter 
($T=40$ MeV and $Y_e=0.4$, lower panel). 
We observe that hyperons do not appear at all in the NS mixed phase, 
because the onset for the hadron-quark phase transition takes place at densities 
below the hyperon onset.
In SN matter, thermal hyperons appear already at low density, 
but they are negligible.
Hence, we conclude that the quarks suppress the hyperons in both cases. 

\begin{figure}[t] 
\includegraphics[clip,scale=2]{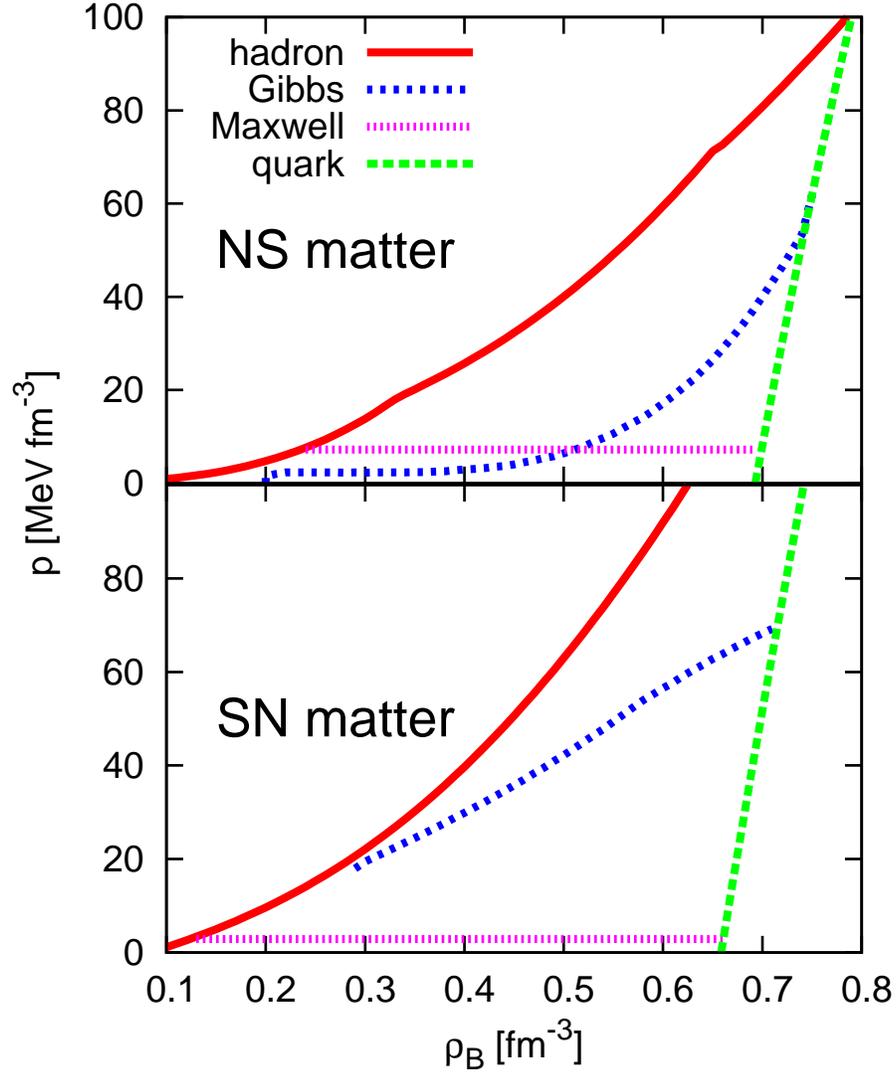}
\caption{
Pressure as a function of the baryon density for NS (upper panel) and 
SN matter (lower panel). 
The red (green) line displays the calculations for 
purely hadronic (quark) matter. 
The blue (pink) curve is the coexistence region when the 
Gibbs (Maxwell) construction is performed.} 
\label{fig:02}
\end{figure}

The resulting EOS $p(\rho_B)$ is displayed in Fig.~\ref{fig:02}, where
the upper (lower) panel displays the calculations performed for NS (SN) matter.
The pure hadronic (quark) phase is represented 
by the solid red (dashed green) curve,
whereas the mixed phase is indicated by the blue (pink) broken line if the
Gibbs (Maxwell) conditions are applied.
We observe that the Gibbs mixed phase spans over a wide range of density, 
around 0.20--0.74 fm$^{-3}$ for NS matter,
and slightly less for SN matter, 0.29--0.71 fm$^{-3}$, 
because finite temperature counteracts the coexistence of phases,
see Ref.~\cite{yasutake09}.
Compared to the EOS with the Gibbs construction, 
the density jump obtained with the Maxwell construction 
appears over a narrower density range in NS matter,
whereas for SN matter the transition takes place at reduced pressure 
and over a wider density range,
which is mainly due to the trapping condition,
as explained in Ref.~\cite{nic06}.

The EOSs discussed above are the fundamental input for 
calculating the stable configurations of compact stars.
For that, we use the well-known hydrostatic equilibrium equations 
of Tolman, Oppenheimer, and Volkov \cite{shapiro}. 
For completeness, 
as in \cite{isen2},
we have attached the Shen EOS \cite{shen98} 
in the low-density regime ($\rho<0.1\;\mathrm{fm}^{-3}$), 
since below this threshold 
clusterization sets in and nuclear matter becomes inhomogeneous. 

\begin{figure}[t] 
\includegraphics[clip,scale=2]{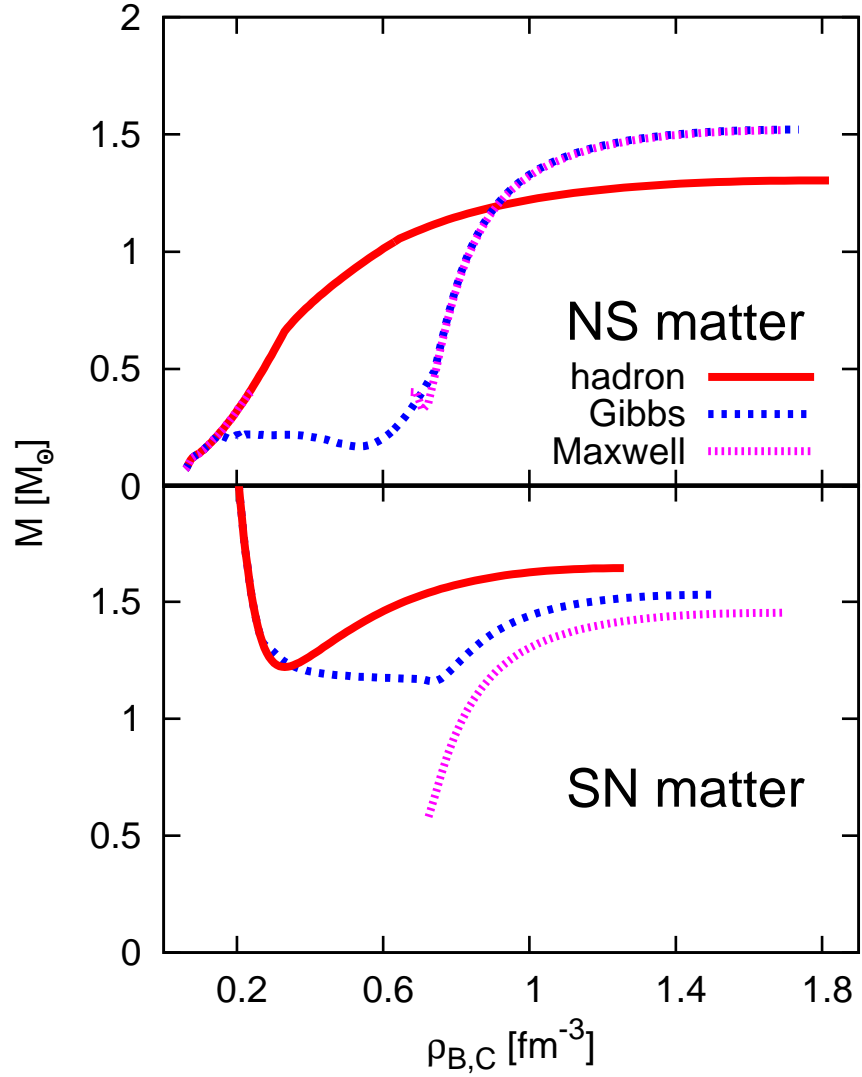} 
\caption{
Gravitational mass 
(in units of the solar mass) 
as a function of the central baryon density $\rho_{B,c}$ 
for NS (upper panel) and SN (lower panel) matter.
The purely hadronic configurations are represented by broken blue lines, 
whereas the red (green) lines represent the configurations of 
hybrid stars obtained applying the Gibbs (Maxwell) construction.}
\label{fig:03}
\end{figure}

In Fig.~\ref{fig:03} we display the gravitational mass
(in units of the solar mass, $\ms = 1.99 \times 10^{33} g$) 
as function of the central baryon number density $\rho_{B,C}$.
The upper (lower) panel shows results for NS (SN) matter, and
the blue (pink) broken lines represent the calculations performed 
with the Gibbs (Maxwell) construction.
For comparison, we also display
the stable configurations in the pure hadronic phase (solid red line).

We find that in NSs the pure hadronic EOS with hyperons
gives rise to a maximum mass smaller than the canonical value, i.e.
1.44 $\ms$,
whereas the hadron-quark phase transition allows to reach maximum
masses slightly above 1.5 $\ms$
(the difference between Gibbs and Maxwell case being very small).
This increase is due to the fact that the hyperon populations 
are suppressed by the existence of quarks, 
as shown in Fig.~\ref{fig:01}.

On the contrary, in the SN case all maximum masses are larger than 1.44 $\ms$,
but here the phase transition lowers the maximum mass 
with respect to the purely hadronic matter,
because trapping reduces the hyperon concentrations in the latter case
compared to NS matter.
Also, we observe that the value of the maximum mass with the Gibbs condition 
is slightly higher than the one with the Maxwell construction,
which is due to the overall higher pressure of the matter in the former case,
see Fig.~\ref{fig:02}.

We also find that the stars with a mass close to the maximum one have quark 
cores.

\section{Conclusions}

We have studied the quark-hadron mixed phase in cold neutron stars 
and hot neutrino-trapped protoneutron stars containing also hyperons,
and compared explicitly the Maxwell and the Gibbs phase transition constructions.

We find that pure quark matter appears in the cores of compact stars 
in any situation
and that the hyperon fractions are nearly completely suppressed 
by the appearance of quarks.
Due to this reason the maximum NS mass is increased by the presence 
of quark matter.
However, the simple MIT bag model used here is not capable
to reach currently observed NS masses of about 1.7 $\ms$ \cite{obs},
and it will be an important task for the future to implement more sophisticated
quark models.

Apart from this challenge,
in this paper we did not consider finite-size effects in the mixed phase. 
In particular, it will be interesting to check the influence
of trapped neutrinos on the pasta structures in future works.

\section*{Acknowledgments}

We are grateful to M.~Baldo, T.~Maruyama, and T.~Tatsumi for their warm 
hospitality and fruitful discussions. 
This study was supported in part by the Grants-in-Aid for the 
Scientific Research from the Ministry of Education, 
Science and Culture of Japan (No. 21105512).
We also acknowledge the support of COMPSTAR,
a research and training program of the European Science Foundation.


\newpage

\end{document}